\documentclass[aps,prl,a4paper,twocolumn,showpacs,floatfix,citeautoscript]{revtex4}
\usepackage{graphicx}
\usepackage{amsfonts}
\usepackage{amsmath}
\usepackage{amssymb}
\usepackage{bm}
\usepackage{dcolumn}
\usepackage{hyperref}
\usepackage[usenames]{color}
\hypersetup{pdfborder=0 0 0,colorlinks=true,citecolor=blue,linkcolor=blue}
\begin{document}

\title{Electronics with correlated oxides: SrVO$_3$/SrTiO$_3$ as a Mott transistor}
\author{Zhicheng Zhong$^1$, Markus Wallerberger$^1$, Jan M.\ Tomczak$^1$, Ciro Taranto$^1$, Nicolaus Parragh$^2$, Alessandro Toschi$^1$, Giorgio Sangiovanni$^2$, and Karsten Held$^1$}
\affiliation{$^1$ Institute of Solid State Physics, Vienna University of Technology, A-1040 Vienna, Austria\\
$^2$ Universit\"at W\"urzburg, Am Hubland, D-97074 W\"urzburg, Germany}
\date{\today}

\begin{abstract} 
We employ density functional theory plus dynamical mean field theory and identify the physical origin of why two layers of SrVO$_3$
on a SrTiO$_3$ substrate are insulating: the thin film geometry lifts the orbital degeneracy which in turn triggers a Mott-Hubbard transition.
 Two layers of SrVO$_3$ are just at the verge of a Mott-Hubbard transition and hence ideally suited for technological applications of the Mott-Hubbard transition: the heterostructure is highly sensitive to  strain, electric field, and temperature. A gate voltage can also turn the insulator into a metal, so that a transistor with ideal on-off (metal-insulator) switching properties is realized.
\end{abstract}

\maketitle

In the last years, there has been tremendous experimental progress to grow oxide heterostructures atomic-layer by atomic-layer, brought about by modern deposition techniques such as molecular beam epitaxy and pulsed laser deposition. A key experiment has been the discovery that a two dimensional electron gas (2DEG) with high mobility is created at the interface of two band insulators, LaAlO$_3$ and SrTiO$_3$\cite{Ohtomo:nat04}. This raised the hope that oxide heterostructures might substitute conventional semiconductor electronics one day, at least for specific applications \cite{Mannhart:sci10,Zubko}. Oxide electronics is however still in its infancy compared to the matured field of silicon electronics. Particularly promising are transistors at the scale of 2nm \cite{Cen:sc09}, solar-cells \cite{Assmann:prl13,Guo13,Liang:sr13}, and the possibility to generate spin-polarized currents \cite{Reyren:prl12}. Last but not least, there is high hope that strong electronic correlations make a difference to conventional semiconductors and give rise to new phenomena \cite{Okamoto:nat04,Okamoto:prb11,Jiang2012}.

However, many oxide heterostructures, including the LaAlO$_3$/SrTiO$_3$ prototype, actually do not show strong electronic correlations. Since electronic correlations are weak, bandstructure calculations on the basis of density functional theory (DFT), e.g., within the local density approximation (LDA)\cite{Jones:rmp89}, or even a tight binding modelling \cite{Zhong:prb13} are sufficient:
Such calculations well reproduce or predict experiment, e.g., angular resolved photoemission spectra \cite{Yoshimatsu:sci11,Santander:nat11,Wang:arxiv13}.
A heterostructure where electronic correlations do play a decisive role is, on the other hand, SrVO$_3$/SrTiO$_3$. In the bulk, SrVO$_3$ is a correlated metal with a moderate
renormalization $\sim 2$ of the bandwidth \cite{Sekiyama:prl04, Maiti:prb06,Takizawa:prb09} and a kink in the energy-momentum dispersion \cite{Byczuk:natp07,Nekrasov:prb06,Aizaki:prl12}. SrVO$_3$ has been widely employed \cite{Nekrasov:prb06,Sekiyama:prl04,Pavarini:prl04,Liebsch:prl03,Footnote1} as a testbed material for LDA+dynamical mean field theory (DMFT) calculations \cite{Kotliar:rmp06,Held:ap07,NoteDMFTlayer,Lechermann13}.
Quite surprisingly, recent experiments \cite{Yoshimatsu:prl10} have found that two layers of SrVO$_3$ grown on a SrTiO$_3$ substrate are insulating instead \cite{MgOAl}. On the basis of the one band Hubbard model it has been argued \cite{Yoshimatsu:prl10} that the reduced bandwidth of the thin film is responsible for the Mott insulating state.

In this letter, we present realistic DFT+DMFT calculations and pin point the origin of the insulator to the orbital splitting of the orbitals, caused by the reduced symmetry of the ultrathin film. The reduced bandwidth 
and the enhanced Coulomb interaction of the thin film do {\em not} play the key role. Our calculations demonstrate the high sensitivity of SrVO$_3$ films. A Mott-Hubbard metal-insulator transition can be triggered by small changes of temperature, (uniaxial) pressure, a capping layer, and an electric field. This makes the SrVO$_3$/SrTiO$_3$ heterostructure most promising for applications as sensors or as a Mott transistor with a gate voltage controlling the Mott-Hubbard transition.

{\em Method.} We perform DFT+DMFT calculations for two layers of SrVO$_3$ 
on a substrate given by four unit cells of SrTiO$_3$ and a sufficiently thick vacuum of 10{\AA} along the $z$ direction. We fix the in-plane ($x$-$y$ plane) lattice constant to the calculated equilibrium bulk value of the substrate $a_{\rm SrTiO_3} = 3.92$\AA, and optimize the internal coordinates. The DFT calculations are performed using the all-electron full potential augmented plane-wave method of the  Wien2K package \cite{WIEN2K} with the generalized gradient approximation (GGA) potential \cite{PerdewPRL96} and a 10$\times$10$\times$1 $k$-point grid.

The DFT states near the Fermi-level are mainly of Vanadium t$_{2g}$ orbital characters, which are well-localized and exhibit strong correlations
beyond DFT and DFT+U. To properly include the correlation effects, we first perform a Wannier projection onto maximally localized \cite{Mostofi2008685} $t_{2g}$ orbitals, using the wien2wannier package \cite{Kune20101888}. We supplement this t$_{2g}$ Hamiltonian constructed from DFT by the local Kanamori Coulomb interaction, given by the
intra-orbital interaction $U$, the inter-orbital (averaged) interaction $U'$ and the Hund's exchange and pair-hopping $J$; for the Hamiltonian see \cite{Parragh:prb12}. For $U'$ we take the calculated constrained LDA value
  $U'=3.55\,$eV \cite{Sekiyama:prl04,Nekrasov:prb06} and for the Hund's exchange a reasonable value for early transition metal oxides:
 $J=0.75\,$eV. Note that constrained LDA tends give too large 
estimates for $J$, see the discussion
in \cite{Held:ap07}; $U=U'+2J$ by symmetry. For the DMFT calculations we use the 
w2dynamics package\cite{Parragh:prb12}, which is an implementation
of  continuous-time quantum Monte Carlo (CT-QMC)
in the hybridization expansion \cite{Gull:rmp11}. We employ the maximum entropy method  \cite{Jarrell:phyreport96} for the  analytical continuation of the spectra to real frequencies. All DMFT calculations are at room temperature if not stated otherwise \cite{Lechermann13}. 

\begin{figure}[tb]
\includegraphics[width=\columnwidth]{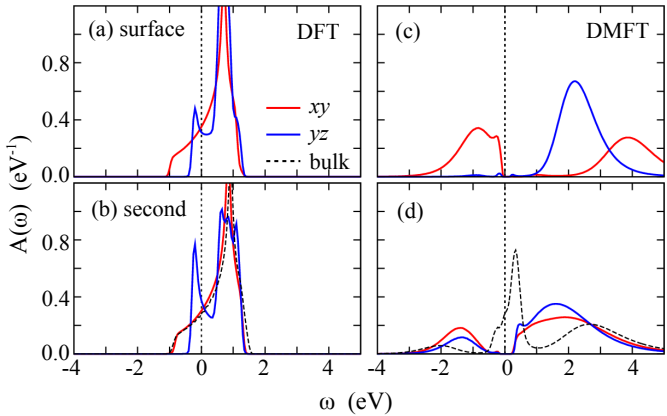}
\caption{Layer-resolved spectral function of two SrVO$_3$ films grown on SrTiO$_3$ in DFT (left) and DFT+DMFT at room temperature (right). The dashed line in the second layer shows the SrVO$_3$ bulk DFT and DFT+DMFT spectrum.}
\label{Fig1}
\end{figure}

{\it DFT+DMFT results for SVO/STO.} Fig.\ \ref{Fig1} shows the DFT and 
DFT+DMFT spectrum for two layers of SrVO$_3$ on a SrTiO$_3$ substrate,
as well as bulk SrVO$_3$ for comparison. Without electronic correlations, i.e.,
within DFT, the $xy$-states are showing (almost) the same spectrum as for the bulk. These 
states have their orbital lobes within the $xy$-plane and can be well modeled with a nearest neighbor hopping that is only in-plane \cite{Zhong:prb13}. Hence, the confinement along the $z$-axis has little effect.
The $yz$-states (and by symmetry the $xz$-states) have a nearest neighbor hopping along the $z$-axis, which is cut-off
by the vacuum and the insulating SrTiO$_3$ substrate \cite{Zhong:prb13}. As a consequence these states become more one-dimensional 
($y$-axis hopping only) and the $yz$-bandwidth is reduced. The $yz$-bands are also pushed up in energy since breaking the cubic symmetry leads to a crystal field splitting $\Delta$=0.18eV between $xy$ and $yz$ ($xz$) states.

This lifting of the orbital degeneracy has dramatic consequences when electronic correlations are taken into account. Indeed it is the physical origin of why  thin SrVO$_3$ films are insulating. The right panels of Fig.\ \ref{Fig1} show the DFT+DMFT spectra, the DMFT self energies are given in the Supplementary Material. For the topmost (surface) layer, we see that 
electronic correlations further push the $yz$- (and $xz$-) states 
up in energy, they are essentially depopulated. That means on the other hand
that the $xy$-states are half-filled. Because of this effective one band situation and the relatively large intra-orbital Coulomb interaction $U$, the
 $xy$-states are Mott-Hubbard split into an upper and lower Hubbard band. The SrVO$_3$ film is a Mott insulator. With the
surface layer being insulating, also the second layer becomes a Mott insulator, albeit here the difference between $xy$ and $yz$ population is much less pronounced. Let us emphasize that due to the DMFT self-consistency also a more insulating second layer feeds back into an even more insulating surface layer.

\begin{figure}[tb]
\includegraphics[width=\columnwidth]{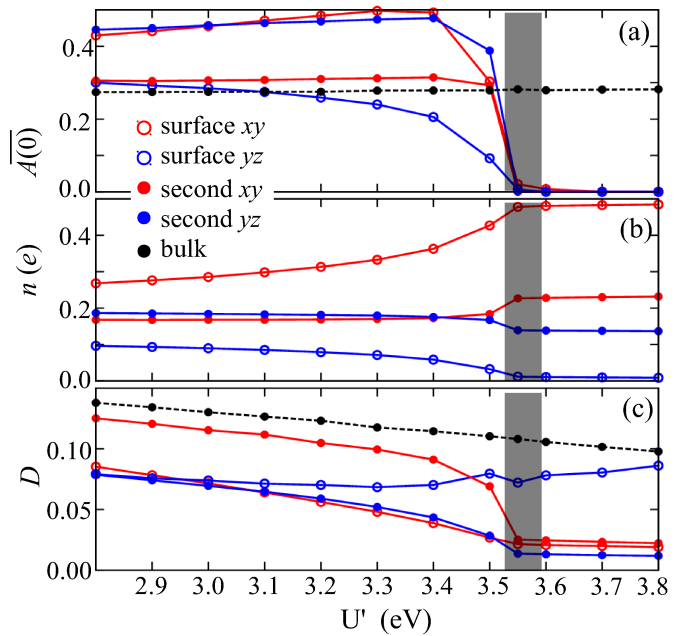}
\caption{Mott-Hubbard transition as a function of interaction $U'$; the grey shaded region indicates the estimated values of $U'$ \cite{Footnote2}. (a): Spectral function around the Fermi level $\overline A(\omega=0)$ -- layer and orbitally resolved. (b): Orbital occupation. (c): Intra-orbital  double occupation.}
\label{Fig2}
\end{figure}

This mutual influence can be inferred from Fig.\ \ref{Fig2} (a), which shows that both layers
get insulating at the same interaction strength. Here, $\overline{A(0)}\equiv \beta G(\tau$=$\beta/2)/ \pi$ can be calculated directly from the CT-QMC data without analytic continuation ($\beta=1/T$ is the inverse temperature); $\overline{A(0)}$ is the spectral function around the Fermi level averaged over a frequency interval $\sim T$.
The orbital occupations  $n_{i \alpha \sigma}= \langle c^{\dagger}_{i \alpha\sigma} c^{\phantom{\dagger}}_{i \alpha\sigma} \rangle$ of the two layers $i$ and orbitals $\alpha$
 in Fig.\ \ref{Fig2} (b) reflect what we have already qualitatively inferred from the spectra in
 Fig.\ \ref{Fig1}: the surface layer becomes fully orbitally polarized, whereas the second layer shows only small differences in the orbital occupation. 
We did not observe any spin ordering in DMFT.
 Fig.\ \ref{Fig2} (c) shows how the normalized
double occupation $ D_{i \alpha}=\langle c^{\dagger}_{i \alpha\uparrow} c^{\phantom{\dagger}}_{i \alpha\uparrow} c^{\dagger}_{i \alpha\downarrow} c^{\phantom{\dagger}}_{i \alpha\downarrow}\rangle /(n_{i \alpha \uparrow}n_{i \alpha \downarrow})$
drops at the Mott-Hubbard transition. The constant behavior of the surface layer $yz$ orbital is simply due to the normalization (numerator and denominator become extremely small).

Altogether, Figs.\ \ref{Fig1} and \ref{Fig2} show the typical behavior of
a Mott-Hubbard transition in a multi-orbital system which is controlled by the ratio interaction-to-bandwidth {\em and} the orbital splitting. The latter
is further enhanced by electronic correlations and is crucial for the
doping-controlled Mott-transition in, e.g., (Cr$_x$V$_{1-x}$)$_2$O$_3$ \cite{Keller:prb04,Poteryaev:prb07}. Fig.\ \ref{Fig2} also shows that, while the two-layer SrVO$_3$ film is insulating, it is just on the verge of an insulator-to-metal transition. As we will see below, this makes the SrVO$_3$/SrTiO$_3$ heterostructure prone to small changes of the environment such as changing
temperature, pressure or applying an electric field. Before turning to these interesting aspects, let us however first discuss the physical origin 
behind the dramatic changes from SrVO$_3$ bulk to SrVO$_3$ thin films.

{\em Physical origin of the Mott-Hubbard transition.}
By performing a number of additional calculations we have been able to
identify the orbital symmetry breaking as the physical origin behind the 
dramatic difference between insulating SrVO$_3$ films and metallic SrVO$_3$ bulk.

In Ref.\ \cite{Yoshimatsu:prl10, Liebsch:prl03} the reduction of the bandwidth due to  cutting off the  hopping perpendicular to the thin films (or surfaces)
has been held responsible for the enhanced correlation effect and even the Mott-Hubbard transition. We calculated this effect: the reduction in bandwidth for the $yz$ orbital is 20\% and essentially zero for the $xy$ orbitals. Following this picture, one would expect an enhanced $yz$-orbital  occupation \cite{Liebsch:prl03}, which is clearly opposite to our findings.

Another possibility which also changes the interaction-to-bandwidth ratio is an increased Coulomb interaction. Often it is argued that the reduced screening at the surface enhances the Coulomb interaction and makes the surface layers more insulating. To quantify this effect, we have performed constrained random phase approximation (cRPA) calculations \cite{Footnote2}. This was numerically not possible for the full heterostructure structure, but only for two freestanding SrVO$_3$ layers. The cRPA interaction of the SrVO$_3$  film is on average $\sim$ 10\% larger than for the bulk (see Supplementary). This overestimates however the enhancement of the Coulomb interaction,
since the screening of the SrTiO$_3$ substrate is disregarded in our simplified calculation.
As the screening stems from a large energy window in which the
differences between SrVO$_3$ and SrTiO$_3$ are less relevant,
a rough estimate is hence that the actual heterostructure has a 5$\%$ larger interaction strength than the bulk (vacuum on one side instead of two should roughly halve the effect).

Even when combining the reduced bandwidth and enhanced Coulomb interaction, this effect is by far insufficient to make SrVO$_3$ insulating: our DFT+DMFT calculations (not shown) still give a metallic phase for bulk SrVO$_3$ if
the ratio interaction-to-bandwidth is overall increased by 70\%.

Instead, the key for the insulating nature of SrVO$_3$ films is the orbital symmetry breaking, given by the crystal field splitting
and also the different bandwidths of the $xy$ and $yz$ ($xz$) orbitals. The combination of both results already at the DFT level in a 
slightly different orbital occupation. Electronic correlations
largely amplify this orbital polarization, see Fig.\ \ref{Fig2}. With the depopulation of
the $yz$ states, the surface  SrVO$_3$ layer 
effectively becomes a one-($xy$-)band system; and such a one-band system is Mott-insulating already 
at a much smaller 
interaction strength.
Let us note that the lifting of the orbital symmetry was considered in  Ref.\ \cite{Yoshimatsu:prl10} 
as a possible source for the observed deviations  between theory and experiment and as a means to reduce the critical $U'$. But it was not regarded as the primary origin of the Mott-Hubbard transition.

{\em Surface shift of the lower Hubbard band.}
Let us also emphasize the peculiar differences between surface and second layer 
in Fig.\ \ref{Fig1}. Besides the difference in occupation already discussed, also the position of the lower Hubbard band is shifted upwards by 0.5~eV  in the surface layer; its weight and sharpness are enhanced. This effect might actually explain the disagreement between photoemission spectroscopy (PES) and DFT+DMFT \cite{Sekiyama:prl04,Taranto:prb13} regarding the position of the lower Hubbard band. As PES is surface sensitive, the surface layer will contribute strongly to the PES signal. In the Supplementary Material, we confirm that shift and sharpening are a general trend also observed for more (four)  SrVO$_3$ layers, and simulate PES spectra for different penetration depths. Including this surface effect leads to a better agreement with bulk PES (see Supplementary Material). A 20\% upwards shift and 10\% narrowing of the lower Hubbard band in the surface layer has also been reported experimentally \cite{Laverock:prl13}.
 Let us note in passing that there is also a layer-dependence of the quasiparticle weight in the metallic  (more layer) case  \cite{Okamoto:prb11}.

\begin{figure}[t!]
\includegraphics[width=\columnwidth]{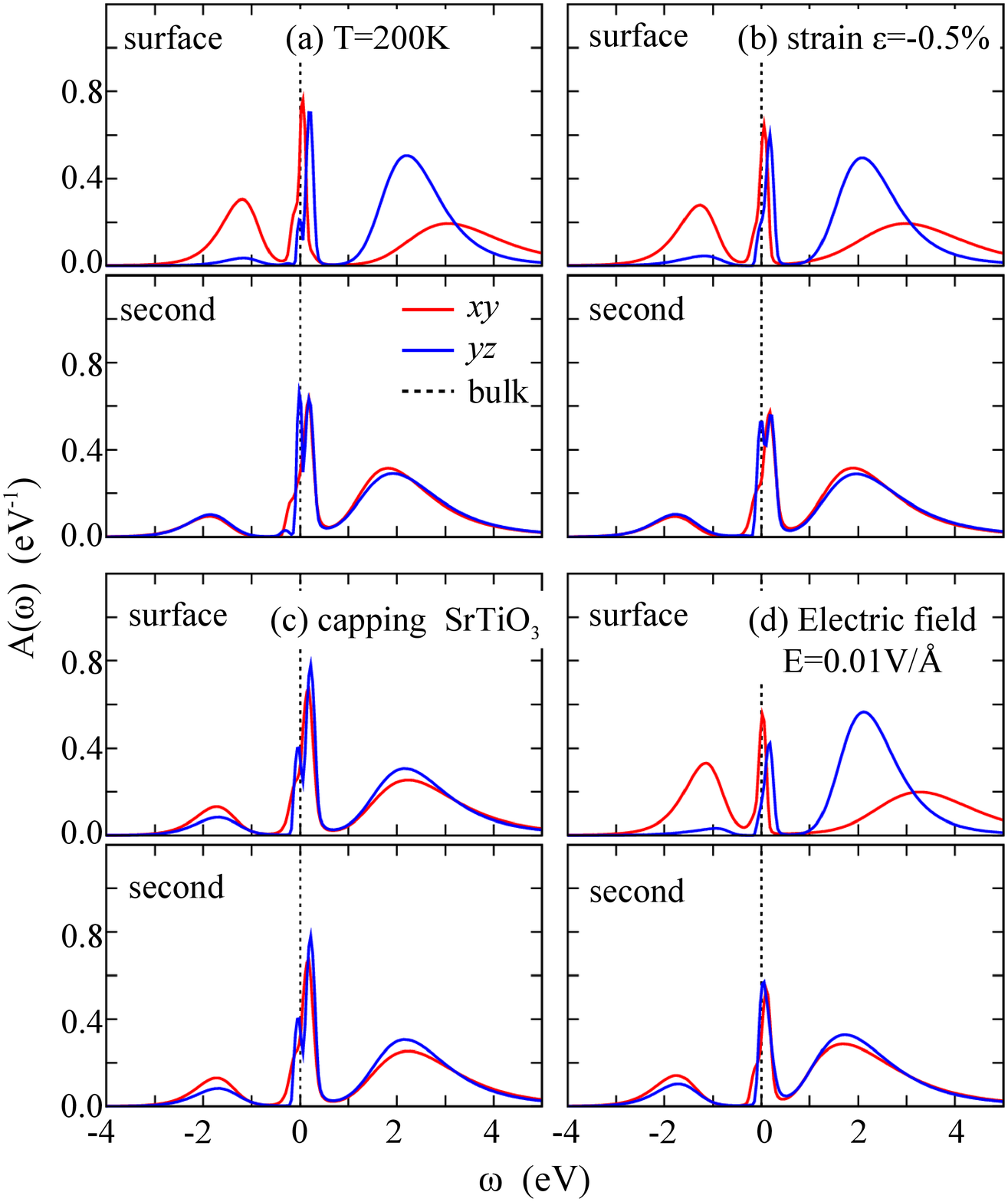}
\caption{The insulating SrVO$_3$/SrTiO$_3$ heterostructure is at the edge of a Mott-Hubbard transition and hence prone to small changes of the environment. (a) Decreasing temperature, (b) a compressive strain, (c) one capping layer of SrTiO$_3$ and (d) an electric field (or gate voltage) switch SrVO$_3$/SrTiO$_3$ from an insulator to a metal. The Figure shows the spectral functions $A(\omega)$ which in all four cases develops a peak at the Fermi level in both SrVO$_3$ layers.}
\label{Fig3}
\end{figure}

{\em Application as a sensor.}
There has been a long quest to make use of strongly correlated electron systems and their huge responses upon small changes of the environment. As we have seen above, two layers of SrVO$_3$ are insulating but on the verge of an insulator-to-metal transition. Hence this heterostructure is highly susceptible to
the most dramatic electronic response possible: from a perfect insulator to a perfect metal. We have identified several mechanisms that will trigger this 
phase transition:\\
(i) Due to its higher entropy, the insulating state is more stable
at higher temperatures. We have tested this hypothesis by additional calculations at 200K: Fig.\ \ref{Fig3} (a) confirms that
the heterostructure switches to a metal.\\
(ii)  There is a sensitivity to pressure (which increases the bandwidth and hence controls the ratio of interaction-to-bandwidth) and uniaxial strain (which controls the important splitting between the $xy$- and $yz$-orbital). In Fig.\ \ref{Fig3} (b) we test 
the effect of the latter by applying a small compressive strain, i.e., decreasing the in-plane lattice constant by 0.5\%. We see that compressive strain turns the SrVO$_3$ layers metallic.\\
(iii) Molecules on the surface also influence the 
electrical behavior, and the heterostructure might be employed as a chemical sensor. We have not investigated the myriad of possibilities. However,
one capping layer of SrTiO$_3$ on top of the SrVO$_3$ layers switches the oxide heterostructure metallic, see Fig.\ \ref{Fig3} (c). An insulating  SrTiO$_3$ capping layer might also serve to protect the 
 SrVO$_3$ layers. In this case, the heterostructure is metallic but can be turned  insulating upon increasing temperature or by applying a tensile strain.\\
(iv) Fig.\ \ref{Fig3} (d) shows that also an external electric field of $0.01$ V/{\AA} triggers the insulator-to-metal transition in the SrVO$_3$/SrTiO$_3$ heterostructure. This corresponds to a gate voltage of 0.08V across the two SrVO$_3$ layers.

Since the Mott-Hubbard transition is an abrupt phase transition, 
it is ideally suited to provide a clear
 {\em on} (metal) and {\em off} (insulator) signal. This signal would indicate that a certain temperature, strain or electric field has been exceeded. On the other hand one can employ the electric field effect to counterbalance the external parameters. This way also quantitative measurements will be possible.

{\em Application as Mott transistor.} 
Thanks to the electric field (or gate voltage) effect, the SrVO$_3$/SrTiO$_3$ heterostructure can also serve as a Mott transistor in oxide electronics.
In this respect, {\em on} (metal) and {\em off} (insulator)
are much better separated than in semiconductor transistors,
where the change in conductivity  with applied gate voltage 
is much more gradual than the abrupt
Mott-Hubbard transition.
In this context, note the considerable experimental efforts to destabilize the (Mott or Peierls) insulating state of bulk VO$_2$ by an electric field \cite{Nakano:nat12,footnote3}.
Recent reports of a field-controlled metal-insulator transition using electrolyte gating \cite{Nakano:nat12} have however been put into question as being caused by field-induced oxygen vacancies \cite{Jeong:sci13}.
The SrVO$_3$/SrTiO$_3$ heterostructure is more promising and better suited for actual electronic devices.

{\em Conclusion.}
We have shown that the physical origin of the insulating nature of
two layers of SrVO$_3$ on a SrTiO$_3$ substrate is the orbital symmetry breaking and polarization, not the change of the bandwidth as previously assumed. The SrVO$_3$/SrTiO$_3$ heterostructure is at the tipping point of a Mott-Hubbard insulator-to-metal transition. Small changes in temperature, strain, electric field
or a capping layers will tip the heterostructure to a metallic behavior.
This can be technologically employed in form of sensors or as a Mott transistor, which actually might form the basis for future oxide electronics.

{\em Acknowledgments.}
 We acknowledge financial support from the SFB ViCoM (ZZ, Austrian Science Fund project ID F4103-N13) through the DFG research unit FOR 1346 (GS) and the European Research Council under the European Union's Seventh Framework Program (FP/2007-2013)/ERC through grant agreement n.\ 306447 (JT,KH).
 Calculations have been done on the Vienna Scientific Cluster~(VSC).

\begin{thebibliography}{40}
\expandafter\ifx\csname natexlab\endcsname\relax\def\natexlab#1{#1}\fi
\expandafter\ifx\csname bibnamefont\endcsname\relax
  \def\bibnamefont#1{#1}\fi
\expandafter\ifx\csname bibfnamefont\endcsname\relax
  \def\bibfnamefont#1{#1}\fi
\expandafter\ifx\csname citenamefont\endcsname\relax
  \def\citenamefont#1{#1}\fi
\expandafter\ifx\csname url\endcsname\relax
  \def\url#1{\texttt{#1}}\fi
\expandafter\ifx\csname urlprefix\endcsname\relax\def\urlprefix{URL }\fi
\providecommand{\bibinfo}[2]{#2}
\providecommand{\eprint}[2][]{\url{#2}}

\bibitem[{\citenamefont{Ohtomo and Hwang}(2004)}]{Ohtomo:nat04}
\bibinfo{author}{\bibfnamefont{A.}~\bibnamefont{Ohtomo}} \bibnamefont{and}
  \bibinfo{author}{\bibfnamefont{H.~Y.} \bibnamefont{Hwang}},
  \bibinfo{journal}{Nature} \textbf{\bibinfo{volume}{427}},
  \bibinfo{pages}{423} (\bibinfo{year}{2004}).

\bibitem[{\citenamefont{Mannhart and Schlom}(2010)}]{Mannhart:sci10}
\bibinfo{author}{\bibfnamefont{J.}~\bibnamefont{Mannhart}} \bibnamefont{and}
  \bibinfo{author}{\bibfnamefont{D.~G.} \bibnamefont{Schlom}},
  \bibinfo{journal}{Science} \textbf{\bibinfo{volume}{327}},
  \bibinfo{pages}{1607} (\bibinfo{year}{2010}).

\bibitem[{\citenamefont{Zubko et~al.}(2011)\citenamefont{Zubko, Gariglio,
  Gabay, Ghosez, and Triscone}}]{Zubko}
\bibinfo{author}{\bibfnamefont{P.}~\bibnamefont{Zubko}},
  \bibinfo{author}{\bibfnamefont{S.}~\bibnamefont{Gariglio}},
  \bibinfo{author}{\bibfnamefont{M.}~\bibnamefont{Gabay}},
  \bibinfo{author}{\bibfnamefont{P.}~\bibnamefont{Ghosez}}, \bibnamefont{and}
  \bibinfo{author}{\bibfnamefont{J.-M.} \bibnamefont{Triscone}},
  \bibinfo{journal}{Annu. Rev. Condens.Matter Phys.}
  \textbf{\bibinfo{volume}{2}}, \bibinfo{pages}{141} (\bibinfo{year}{2011}).

\bibitem[{\citenamefont{Cen et~al.}(2009)\citenamefont{Cen, Thiel, Mannhart,
  and Levy}}]{Cen:sc09}
\bibinfo{author}{\bibfnamefont{C.}~\bibnamefont{Cen}},
  \bibinfo{author}{\bibfnamefont{S.}~\bibnamefont{Thiel}},
  \bibinfo{author}{\bibfnamefont{J.}~\bibnamefont{Mannhart}}, \bibnamefont{and}
  \bibinfo{author}{\bibfnamefont{J.}~\bibnamefont{Levy}},
  \bibinfo{journal}{Science} \textbf{\bibinfo{volume}{323}},
  \bibinfo{pages}{1026} (\bibinfo{year}{2009}).

\bibitem[{\citenamefont{Assmann et~al.}(2013)\citenamefont{Assmann, Blaha,
  Laskowski, Held, Okamoto, and Sangiovanni}}]{Assmann:prl13}
\bibinfo{author}{\bibfnamefont{E.}~\bibnamefont{Assmann}},
  \bibinfo{author}{\bibfnamefont{P.}~\bibnamefont{Blaha}},
  \bibinfo{author}{\bibfnamefont{R.}~\bibnamefont{Laskowski}},
  \bibinfo{author}{\bibfnamefont{K.}~\bibnamefont{Held}},
  \bibinfo{author}{\bibfnamefont{S.}~\bibnamefont{Okamoto}}, \bibnamefont{and}
  \bibinfo{author}{\bibfnamefont{G.}~\bibnamefont{Sangiovanni}},
  \bibinfo{journal}{Phys. Rev. Lett.} \textbf{\bibinfo{volume}{110}},
  \bibinfo{pages}{078701} (\bibinfo{year}{2013}).

\bibitem{Guo13}
 H.-Z. Guo, L. Gu, Z.-Z. Yang, S.-F. Wang, G.-S. Fu, L. Wang, K.-J. Jin, H.-B. Lu, C. Wang, C. Ge, M. He, and G.-Z. Yang, Euro. Phys. Lett. {\bf 103}, 47006 (2013).

\bibitem[{\citenamefont{Liang et~al.}(2013)\citenamefont{Liang, Cheng, Zhai,
  Pan, Guo, Zhao, Zhang, Li, Zhang, Wang et~al.}}]{Liang:sr13}
\bibinfo{author}{\bibfnamefont{H.}~\bibnamefont{Liang}},
  \bibinfo{author}{\bibfnamefont{L.}~\bibnamefont{Cheng}},
  \bibinfo{author}{\bibfnamefont{X.}~\bibnamefont{Zhai}},
  \bibinfo{author}{\bibfnamefont{N.}~\bibnamefont{Pan}},
  \bibinfo{author}{\bibfnamefont{H.}~\bibnamefont{Guo}},
  \bibinfo{author}{\bibfnamefont{J.}~\bibnamefont{Zhao}},
  \bibinfo{author}{\bibfnamefont{H.}~\bibnamefont{Zhang}},
  \bibinfo{author}{\bibfnamefont{L.}~\bibnamefont{Li}},
  \bibinfo{author}{\bibfnamefont{X.}~\bibnamefont{Zhang}},
  \bibinfo{author}{\bibfnamefont{X.}~\bibnamefont{Wang}}, \bibnamefont{et~al.},
  \bibinfo{journal}{Scientific Reports} \textbf{\bibinfo{volume}{3}},
  \bibinfo{pages}{1975} (\bibinfo{year}{2013}).

\bibitem[{\citenamefont{Reyren et~al.}(2012)\citenamefont{Reyren, Bibes, Lesne,
  George, Deranlot, Collin, Barth\'el\'emy, and Jaffr\`es}}]{Reyren:prl12}
\bibinfo{author}{\bibfnamefont{N.}~\bibnamefont{Reyren}},
  \bibinfo{author}{\bibfnamefont{M.}~\bibnamefont{Bibes}},
  \bibinfo{author}{\bibfnamefont{E.}~\bibnamefont{Lesne}},
  \bibinfo{author}{\bibfnamefont{J.-M.} \bibnamefont{George}},
  \bibinfo{author}{\bibfnamefont{C.}~\bibnamefont{Deranlot}},
  \bibinfo{author}{\bibfnamefont{S.}~\bibnamefont{Collin}},
  \bibinfo{author}{\bibfnamefont{A.}~\bibnamefont{Barth\'el\'emy}},
  \bibnamefont{and}
  \bibinfo{author}{\bibfnamefont{H.}~\bibnamefont{Jaffr\`es}},
  \bibinfo{journal}{Phys. Rev. Lett.} \textbf{\bibinfo{volume}{108}},
  \bibinfo{pages}{186802} (\bibinfo{year}{2012}).

\bibitem[{\citenamefont{Okamoto and Millis}(2004)}]{Okamoto:nat04}
\bibinfo{author}{\bibfnamefont{S.}~\bibnamefont{Okamoto}} \bibnamefont{and}
  \bibinfo{author}{\bibfnamefont{A.~J.} \bibnamefont{Millis}},
  \bibinfo{journal}{Nature} \textbf{\bibinfo{volume}{428}},
  \bibinfo{pages}{630} (\bibinfo{year}{2004}).

\bibitem{Okamoto:prb11}
S. Okamoto, Phys. Rev. B {\bf 84}, 201305(R) (2011).
\bibitem{Jiang2012}
M. Jiang, G. G. Batrouni, and R. T. Scalettar,
Phys. Rev. B 86, 195117 (2012). 
\bibitem[{\citenamefont{Jones and Gunnarsson}(1989)}]{Jones:rmp89}
\bibinfo{author}{\bibfnamefont{R.~O.} \bibnamefont{Jones}} \bibnamefont{and}
  \bibinfo{author}{\bibfnamefont{O.}~\bibnamefont{Gunnarsson}},
  \bibinfo{journal}{Rev. Mod. Phys.} \textbf{\bibinfo{volume}{61}},
  \bibinfo{pages}{689} (\bibinfo{year}{1989}).

\bibitem[{\citenamefont{Zhong et~al.}(2013)\citenamefont{Zhong, Zhang, and
  Held}}]{Zhong:prb13}
\bibinfo{author}{\bibfnamefont{Z.}~\bibnamefont{Zhong}},
  \bibinfo{author}{\bibfnamefont{Q.}~\bibnamefont{Zhang}}, \bibnamefont{and}
  \bibinfo{author}{\bibfnamefont{K.}~\bibnamefont{Held}},
  \bibinfo{journal}{Phys. Rev. B} \textbf{\bibinfo{volume}{88}},
  \bibinfo{pages}{125401} (\bibinfo{year}{2013}).

\bibitem[{\citenamefont{Yoshimatsu et~al.}(2011)\citenamefont{Yoshimatsu,
  Horiba, Kumigashira, Yoshida, Fujimori, and Oshima}}]{Yoshimatsu:sci11}
\bibinfo{author}{\bibfnamefont{K.}~\bibnamefont{Yoshimatsu}},
  \bibinfo{author}{\bibfnamefont{K.}~\bibnamefont{Horiba}},
  \bibinfo{author}{\bibfnamefont{H.}~\bibnamefont{Kumigashira}},
  \bibinfo{author}{\bibfnamefont{T.}~\bibnamefont{Yoshida}},
  \bibinfo{author}{\bibfnamefont{A.}~\bibnamefont{Fujimori}}, \bibnamefont{and}
  \bibinfo{author}{\bibfnamefont{M.}~\bibnamefont{Oshima}},
  \bibinfo{journal}{Science} \textbf{\bibinfo{volume}{333}},
  \bibinfo{pages}{319} (\bibinfo{year}{2011}).

\bibitem[{\citenamefont{Santander-Syro
  et~al.}(2011)\citenamefont{Santander-Syro, Copie, Kondo, Fortuna, Pailhès,
  Weht, Qiu, Bertran, Nicolaou, Taleb-Ibrahimi et~al.}}]{Santander:nat11}
\bibinfo{author}{\bibfnamefont{A.~F.} \bibnamefont{Santander-Syro}},
  \bibinfo{author}{\bibfnamefont{O.}~\bibnamefont{Copie}},
  \bibinfo{author}{\bibfnamefont{T.}~\bibnamefont{Kondo}},
  \bibinfo{author}{\bibfnamefont{F.}~\bibnamefont{Fortuna}},
  \bibinfo{author}{\bibfnamefont{S.}~\bibnamefont{Pailhès}},
  \bibinfo{author}{\bibfnamefont{R.}~\bibnamefont{Weht}},
  \bibinfo{author}{\bibfnamefont{X.~G.} \bibnamefont{Qiu}},
  \bibinfo{author}{\bibfnamefont{F.}~\bibnamefont{Bertran}},
  \bibinfo{author}{\bibfnamefont{A.}~\bibnamefont{Nicolaou}},
  \bibinfo{author}{\bibfnamefont{A.}~\bibnamefont{Taleb-Ibrahimi}},
  \bibnamefont{et~al.}, \bibinfo{journal}{Nature}
  \textbf{\bibinfo{volume}{469}}, \bibinfo{pages}{189} (\bibinfo{year}{2011}).

\bibitem[{\citenamefont{Wang et~al.}(2013)\citenamefont{Wang, Zhong, Hao,
  Gerhold, Stoger, Schmid, Sanchez-Barriga, Varykhalov, Franchini, Held
  et~al.}}]{Wang:arxiv13}
\bibinfo{author}{\bibfnamefont{Z.}~\bibnamefont{Wang}},
  \bibinfo{author}{\bibfnamefont{Z.}~\bibnamefont{Zhong}},
  \bibinfo{author}{\bibfnamefont{X.}~\bibnamefont{Hao}},
  \bibinfo{author}{\bibfnamefont{S.}~\bibnamefont{Gerhold}},
  \bibinfo{author}{\bibfnamefont{B.}~\bibnamefont{Stoger}},
  \bibinfo{author}{\bibfnamefont{M.}~\bibnamefont{Schmid}},
  \bibinfo{author}{\bibfnamefont{J.}~\bibnamefont{Sanchez-Barriga}},
  \bibinfo{author}{\bibfnamefont{A.}~\bibnamefont{Varykhalov}},
  \bibinfo{author}{\bibfnamefont{C.}~\bibnamefont{Franchini}},
  \bibinfo{author}{\bibfnamefont{K.}~\bibnamefont{Held}}, \bibnamefont{et~al.},
  \bibinfo{journal}{Arxiv} \textbf{\bibinfo{volume}{1309.7042}}
  (\bibinfo{year}{2013}).

\bibitem[{\citenamefont{Sekiyama et~al.}(2004)\citenamefont{Sekiyama, Fujiwara,
  Imada, Suga, Eisaki, Uchida, Takegahara, Harima, Saitoh, Nekrasov
  et~al.}}]{Sekiyama:prl04}
\bibinfo{author}{\bibfnamefont{A.}~\bibnamefont{Sekiyama}},
  \bibinfo{author}{\bibfnamefont{H.}~\bibnamefont{Fujiwara}},
  \bibinfo{author}{\bibfnamefont{S.}~\bibnamefont{Imada}},
  \bibinfo{author}{\bibfnamefont{S.}~\bibnamefont{Suga}},
  \bibinfo{author}{\bibfnamefont{H.}~\bibnamefont{Eisaki}},
  \bibinfo{author}{\bibfnamefont{S.~I.} \bibnamefont{Uchida}},
  \bibinfo{author}{\bibfnamefont{K.}~\bibnamefont{Takegahara}},
  \bibinfo{author}{\bibfnamefont{H.}~\bibnamefont{Harima}},
  \bibinfo{author}{\bibfnamefont{Y.}~\bibnamefont{Saitoh}},
  \bibinfo{author}{\bibfnamefont{I.~A.} \bibnamefont{Nekrasov}},
  \bibnamefont{et~al.}, \bibinfo{journal}{Phys. Rev. Lett.}
  \textbf{\bibinfo{volume}{93}}, \bibinfo{pages}{156402}
  (\bibinfo{year}{2004}).

\bibitem[{\citenamefont{Maiti et~al.}(2006)\citenamefont{Maiti, Manju, Ray,
  Mahadevan, Inoue, Carbone, and Sarma}}]{Maiti:prb06}
\bibinfo{author}{\bibfnamefont{K.}~\bibnamefont{Maiti}},
  \bibinfo{author}{\bibfnamefont{U.}~\bibnamefont{Manju}},
  \bibinfo{author}{\bibfnamefont{S.}~\bibnamefont{Ray}},
  \bibinfo{author}{\bibfnamefont{P.}~\bibnamefont{Mahadevan}},
  \bibinfo{author}{\bibfnamefont{I.~H.} \bibnamefont{Inoue}},
  \bibinfo{author}{\bibfnamefont{C.}~\bibnamefont{Carbone}}, \bibnamefont{and}
  \bibinfo{author}{\bibfnamefont{D.~D.} \bibnamefont{Sarma}},
  \bibinfo{journal}{Phys. Rev. B} \textbf{\bibinfo{volume}{73}},
  \bibinfo{pages}{052508} (\bibinfo{year}{2006}).

\bibitem[{\citenamefont{Takizawa et~al.}(2009)\citenamefont{Takizawa, Minohara,
  Kumigashira, Toyota, Oshima, Wadati, Yoshida, Fujimori, Lippmaa, Kawasaki
  et~al.}}]{Takizawa:prb09}
\bibinfo{author}{\bibfnamefont{M.}~\bibnamefont{Takizawa}},
  \bibinfo{author}{\bibfnamefont{M.}~\bibnamefont{Minohara}},
  \bibinfo{author}{\bibfnamefont{H.}~\bibnamefont{Kumigashira}},
  \bibinfo{author}{\bibfnamefont{D.}~\bibnamefont{Toyota}},
  \bibinfo{author}{\bibfnamefont{M.}~\bibnamefont{Oshima}},
  \bibinfo{author}{\bibfnamefont{H.}~\bibnamefont{Wadati}},
  \bibinfo{author}{\bibfnamefont{T.}~\bibnamefont{Yoshida}},
  \bibinfo{author}{\bibfnamefont{A.}~\bibnamefont{Fujimori}},
  \bibinfo{author}{\bibfnamefont{M.}~\bibnamefont{Lippmaa}},
  \bibinfo{author}{\bibfnamefont{M.}~\bibnamefont{Kawasaki}},
  \bibnamefont{et~al.}, \bibinfo{journal}{Phys. Rev. B}
  \textbf{\bibinfo{volume}{80}}, \bibinfo{pages}{235104}
  (\bibinfo{year}{2009}).

\bibitem[{\citenamefont{Byczuk et~al.}(2007)\citenamefont{Byczuk, Kollar, Held,
  Yang, Nekrasov, Pruschke, and Vollhardt}}]{Byczuk:natp07}
\bibinfo{author}{\bibfnamefont{K.}~\bibnamefont{Byczuk}},
  \bibinfo{author}{\bibfnamefont{M.}~\bibnamefont{Kollar}},
  \bibinfo{author}{\bibfnamefont{K.}~\bibnamefont{Held}},
  \bibinfo{author}{\bibfnamefont{Y.-F.} \bibnamefont{Yang}},
  \bibinfo{author}{\bibfnamefont{I.~A.} \bibnamefont{Nekrasov}},
  \bibinfo{author}{\bibfnamefont{T.}~\bibnamefont{Pruschke}}, \bibnamefont{and}
  \bibinfo{author}{\bibfnamefont{D.}~\bibnamefont{Vollhardt}},
  \bibinfo{journal}{Nature Physics} \textbf{\bibinfo{volume}{3}},
  \bibinfo{pages}{168} (\bibinfo{year}{2007}).

\bibitem[{\citenamefont{Nekrasov et~al.}(2006)\citenamefont{Nekrasov, Held,
  Keller, Kondakov, Pruschke, Kollar, Andersen, Anisimov, and
  Vollhardt}}]{Nekrasov:prb06}
\bibinfo{author}{\bibfnamefont{I.~A.} \bibnamefont{Nekrasov}},
  \bibinfo{author}{\bibfnamefont{K.}~\bibnamefont{Held}},
  \bibinfo{author}{\bibfnamefont{G.}~\bibnamefont{Keller}},
  \bibinfo{author}{\bibfnamefont{D.~E.} \bibnamefont{Kondakov}},
  \bibinfo{author}{\bibfnamefont{T.}~\bibnamefont{Pruschke}},
  \bibinfo{author}{\bibfnamefont{M.}~\bibnamefont{Kollar}},
  \bibinfo{author}{\bibfnamefont{O.~K.} \bibnamefont{Andersen}},
  \bibinfo{author}{\bibfnamefont{V.~I.} \bibnamefont{Anisimov}},
  \bibnamefont{and}
  \bibinfo{author}{\bibfnamefont{D.}~\bibnamefont{Vollhardt}},
  \bibinfo{journal}{Phys. Rev. B} \textbf{\bibinfo{volume}{73}},
  \bibinfo{pages}{155112} (\bibinfo{year}{2006}).

\bibitem[{\citenamefont{Aizaki et~al.}(2012)\citenamefont{Aizaki, Yoshida,
  Yoshimatsu, Takizawa, Minohara, Ideta, Fujimori, Gupta, Mahadevan, Horiba
  et~al.}}]{Aizaki:prl12}
\bibinfo{author}{\bibfnamefont{S.}~\bibnamefont{Aizaki}},
  \bibinfo{author}{\bibfnamefont{T.}~\bibnamefont{Yoshida}},
  \bibinfo{author}{\bibfnamefont{K.}~\bibnamefont{Yoshimatsu}},
  \bibinfo{author}{\bibfnamefont{M.}~\bibnamefont{Takizawa}},
  \bibinfo{author}{\bibfnamefont{M.}~\bibnamefont{Minohara}},
  \bibinfo{author}{\bibfnamefont{S.}~\bibnamefont{Ideta}},
  \bibinfo{author}{\bibfnamefont{A.}~\bibnamefont{Fujimori}},
  \bibinfo{author}{\bibfnamefont{K.}~\bibnamefont{Gupta}},
  \bibinfo{author}{\bibfnamefont{P.}~\bibnamefont{Mahadevan}},
  \bibinfo{author}{\bibfnamefont{K.}~\bibnamefont{Horiba}},
  \bibnamefont{et~al.}, \bibinfo{journal}{Phys. Rev. Lett.}
  \textbf{\bibinfo{volume}{109}}, \bibinfo{pages}{056401}
  (\bibinfo{year}{2012}).

\bibitem[{\citenamefont{Pavarini et~al.}(2004)\citenamefont{Pavarini, Biermann,
  Poteryaev, Lichtenstein, Georges, and Andersen}}]{Pavarini:prl04}
\bibinfo{author}{\bibfnamefont{E.}~\bibnamefont{Pavarini}},
  \bibinfo{author}{\bibfnamefont{S.}~\bibnamefont{Biermann}},
  \bibinfo{author}{\bibfnamefont{A.}~\bibnamefont{Poteryaev}},
  \bibinfo{author}{\bibfnamefont{A.~I.} \bibnamefont{Lichtenstein}},
  \bibinfo{author}{\bibfnamefont{A.}~\bibnamefont{Georges}}, \bibnamefont{and}
  \bibinfo{author}{\bibfnamefont{O.~K.} \bibnamefont{Andersen}},
  \bibinfo{journal}{Phys. Rev. Lett.} \textbf{\bibinfo{volume}{92}},
  \bibinfo{pages}{176403} (\bibinfo{year}{2004}).

\bibitem[{\citenamefont{Liebsch}(2003)}]{Liebsch:prl03}
\bibinfo{author}{\bibfnamefont{A.}~\bibnamefont{Liebsch}},
  \bibinfo{journal}{Phys. Rev. Lett.} \textbf{\bibinfo{volume}{90}},
  \bibinfo{pages}{096401} (\bibinfo{year}{2003}).

\bibitem[{Foo({\natexlab{a}})}]{Footnote1}
\bibinfo{journal}{As well as for $GW$+DMFT calculations, see Ref.
  \cite{Tomczak:epl12,Taranto:prb13}}.

\bibitem[{\citenamefont{Kotliar et~al.}(2006)\citenamefont{Kotliar, Savrasov,
  Haule, Oudovenko, Parcollet, and Marianetti}}]{Kotliar:rmp06}
\bibinfo{author}{\bibfnamefont{G.}~\bibnamefont{Kotliar}},
  \bibinfo{author}{\bibfnamefont{S.~Y.} \bibnamefont{Savrasov}},
  \bibinfo{author}{\bibfnamefont{K.}~\bibnamefont{Haule}},
  \bibinfo{author}{\bibfnamefont{V.~S.} \bibnamefont{Oudovenko}},
  \bibinfo{author}{\bibfnamefont{O.}~\bibnamefont{Parcollet}},
  \bibnamefont{and} \bibinfo{author}{\bibfnamefont{C.~A.}
  \bibnamefont{Marianetti}}, \bibinfo{journal}{Rev. Mod. Phys.}
  \textbf{\bibinfo{volume}{78}}, \bibinfo{pages}{865} (\bibinfo{year}{2006}).

\bibitem[{\citenamefont{Held}(2007)}]{Held:ap07}
\bibinfo{author}{\bibfnamefont{K.}~\bibnamefont{Held}},
  \bibinfo{journal}{Advances in Physics} \textbf{\bibinfo{volume}{56}},
  \bibinfo{pages}{829} (\bibinfo{year}{2007}).


\bibitem{NoteDMFTlayer} For layer calculations with DMFT
see  M.~Potthoff and W.~Nolting, Phys. Rev. B, {\bf 60},
7834 (1999); J. K. Freericks, \emph{Transport in multilayered nanostructures: the dynamical mean-field theory approach}, Imperial College Press, London, 2006.
For nanoscopic calculations see
S. Florens,  Phys. Rev. Lett. {\bf 99}, 046402 (2007);
S.~Biermann, A.~Georges, A.~Lichtenstein, and
T.~Giamarchi, Phys. Rev. Lett., {\bf 87}, 276405 (2001);
 M.~Snoek, I.~Titvinidze, C.~T\H{o}ke, K.~Byczuk, and
W.~Hofstetter, New J. Phys. {\bf 10}, 093008 (2008);
 A.~Valli, G.~Sangiovanni, O.~Gunnarsson, A.~Toschi,
and K.~Held, Phys. Rev. Lett. {\bf 104}, 246402 (2010).


\bibitem{Lechermann13} For DFT+DMFT calculations of a 
 LaTiO$_3$/SrTiO$_3$ heterostructure \cite{Lechermann13}, where a Mott and a band insulator get metallic at the interface, see F. Lechermann, L. Boehnke, and D. Grieger, Phys. Rev. B {\bf 87}, 241101(R) (2013).

\bibitem[{\citenamefont{Yoshimatsu et~al.}(2010)\citenamefont{Yoshimatsu,
  Okabe, Kumigashira, Okamoto, Aizaki, Fujimori, and
  Oshima}}]{Yoshimatsu:prl10}
\bibinfo{author}{\bibfnamefont{K.}~\bibnamefont{Yoshimatsu}},
  \bibinfo{author}{\bibfnamefont{T.}~\bibnamefont{Okabe}},
  \bibinfo{author}{\bibfnamefont{H.}~\bibnamefont{Kumigashira}},
  \bibinfo{author}{\bibfnamefont{S.}~\bibnamefont{Okamoto}},
  \bibinfo{author}{\bibfnamefont{S.}~\bibnamefont{Aizaki}},
  \bibinfo{author}{\bibfnamefont{A.}~\bibnamefont{Fujimori}}, \bibnamefont{and}
  \bibinfo{author}{\bibfnamefont{M.}~\bibnamefont{Oshima}},
  \bibinfo{journal}{Phys. Rev. Lett.} \textbf{\bibinfo{volume}{104}},
  \bibinfo{pages}{147601} (\bibinfo{year}{2010}).

\bibitem{MgOAl} Weakly correlated surfaces usually stay metallic to a single layer, see e.g.,  R. Arita, Y. Tanida, S. Entani, M. Kiguchi, K. Saiki, and H. Aoki, Phys. Rev. B {\bf 69}, 235423 (2004).

\bibitem[{\citenamefont{Blaha et~al.}(2001)\citenamefont{Blaha, Schwarz,
  Madsen, Kvasnicka, and Luitz}}]{WIEN2K}
\bibinfo{author}{\bibfnamefont{P.}~\bibnamefont{Blaha}},
  \bibinfo{author}{\bibfnamefont{K.}~\bibnamefont{Schwarz}},
  \bibinfo{author}{\bibfnamefont{G.~K.~H.} \bibnamefont{Madsen}},
  \bibinfo{author}{\bibfnamefont{D.}~\bibnamefont{Kvasnicka}},
  \bibnamefont{and} \bibinfo{author}{\bibfnamefont{J.}~\bibnamefont{Luitz}},
  \emph{\bibinfo{title}{WIEN2k, An Augmented Plane Wave + Local Orbitals
  Program for Calculating Crystal Properties}} (\bibinfo{publisher}{Karlheinz
  Schwarz, Techn. Universit\"at Wien, Austria}, \bibinfo{year}{2001}), ISBN
  \bibinfo{isbn}{3-9501031-1-2}.

\bibitem[{\citenamefont{Perdew et~al.}(1996)\citenamefont{Perdew, Burke, and
  Ernzerhof}}]{PerdewPRL96}
\bibinfo{author}{\bibfnamefont{J.~P.} \bibnamefont{Perdew}},
  \bibinfo{author}{\bibfnamefont{K.}~\bibnamefont{Burke}}, \bibnamefont{and}
  \bibinfo{author}{\bibfnamefont{M.}~\bibnamefont{Ernzerhof}},
  \bibinfo{journal}{Phys. Rev. Lett.} \textbf{\bibinfo{volume}{77}},
  \bibinfo{pages}{3865} (\bibinfo{year}{1996}).

\bibitem[{\citenamefont{Mostofi et~al.}(2008)\citenamefont{Mostofi, Yates, Lee,
  Souza, Vanderbilt, and Marzari}}]{Mostofi2008685}
\bibinfo{author}{\bibfnamefont{A.~A.} \bibnamefont{Mostofi}},
  \bibinfo{author}{\bibfnamefont{J.~R.} \bibnamefont{Yates}},
  \bibinfo{author}{\bibfnamefont{Y.-S.} \bibnamefont{Lee}},
  \bibinfo{author}{\bibfnamefont{I.}~\bibnamefont{Souza}},
  \bibinfo{author}{\bibfnamefont{D.}~\bibnamefont{Vanderbilt}},
  \bibnamefont{and} \bibinfo{author}{\bibfnamefont{N.}~\bibnamefont{Marzari}},
  \bibinfo{journal}{Computer Physics Communications}
  \textbf{\bibinfo{volume}{178}}, \bibinfo{pages}{685 } (\bibinfo{year}{2008}),
  ISSN \bibinfo{issn}{0010-4655}.

\bibitem[{\citenamefont{Kune\v{s} et~al.}(2010)\citenamefont{Kuneš, Arita,
  Wissgott, Toschi, Ikeda, and Held}}]{Kune20101888}
\bibinfo{author}{\bibfnamefont{J.}~\bibnamefont{Kune\v{s}}},
  \bibinfo{author}{\bibfnamefont{R.}~\bibnamefont{Arita}},
  \bibinfo{author}{\bibfnamefont{P.}~\bibnamefont{Wissgott}},
  \bibinfo{author}{\bibfnamefont{A.}~\bibnamefont{Toschi}},
  \bibinfo{author}{\bibfnamefont{H.}~\bibnamefont{Ikeda}}, \bibnamefont{and}
  \bibinfo{author}{\bibfnamefont{K.}~\bibnamefont{Held}},
  \bibinfo{journal}{Computer Physics Communications}
  \textbf{\bibinfo{volume}{181}}, \bibinfo{pages}{1888 }
  (\bibinfo{year}{2010}), ISSN \bibinfo{issn}{0010-4655}.

\bibitem[{\citenamefont{Parragh et~al.}(2012)\citenamefont{Parragh, Toschi,
  Held, and Sangiovanni}}]{Parragh:prb12}
\bibinfo{author}{\bibfnamefont{N.}~\bibnamefont{Parragh}},
  \bibinfo{author}{\bibfnamefont{A.}~\bibnamefont{Toschi}},
  \bibinfo{author}{\bibfnamefont{K.}~\bibnamefont{Held}}, \bibnamefont{and}
  \bibinfo{author}{\bibfnamefont{G.}~\bibnamefont{Sangiovanni}},
  \bibinfo{journal}{Phys. Rev. B} \textbf{\bibinfo{volume}{86}},
  \bibinfo{pages}{155158} (\bibinfo{year}{2012}).

\bibitem[{\citenamefont{Gull et~al.}(2011)\citenamefont{Gull, Millis,
  Lichtenstein, Rubtsov, Troyer, and Werner}}]{Gull:rmp11}
\bibinfo{author}{\bibfnamefont{E.}~\bibnamefont{Gull}},
  \bibinfo{author}{\bibfnamefont{A.~J.} \bibnamefont{Millis}},
  \bibinfo{author}{\bibfnamefont{A.~I.} \bibnamefont{Lichtenstein}},
  \bibinfo{author}{\bibfnamefont{A.~N.} \bibnamefont{Rubtsov}},
  \bibinfo{author}{\bibfnamefont{M.}~\bibnamefont{Troyer}}, \bibnamefont{and}
  \bibinfo{author}{\bibfnamefont{P.}~\bibnamefont{Werner}},
  \bibinfo{journal}{Rev. Mod. Phys.} \textbf{\bibinfo{volume}{83}},
  \bibinfo{pages}{349} (\bibinfo{year}{2011}).

\bibitem[{\citenamefont{Jarrell and Gubernatis}(1996)}]{Jarrell:phyreport96}
\bibinfo{author}{\bibfnamefont{M.}~\bibnamefont{Jarrell}} \bibnamefont{and}
  \bibinfo{author}{\bibfnamefont{J.}~\bibnamefont{Gubernatis}},
  \bibinfo{journal}{Physics Reports} \textbf{\bibinfo{volume}{269}},
  \bibinfo{pages}{133} (\bibinfo{year}{1996}).

\bibitem[{\citenamefont{Keller et~al.}(2004)\citenamefont{Keller, Held, Eyert,
  Vollhardt, and Anisimov}}]{Keller:prb04}
\bibinfo{author}{\bibfnamefont{G.}~\bibnamefont{Keller}},
  \bibinfo{author}{\bibfnamefont{K.}~\bibnamefont{Held}},
  \bibinfo{author}{\bibfnamefont{V.}~\bibnamefont{Eyert}},
  \bibinfo{author}{\bibfnamefont{D.}~\bibnamefont{Vollhardt}},
  \bibnamefont{and} \bibinfo{author}{\bibfnamefont{V.~I.}
  \bibnamefont{Anisimov}}, \bibinfo{journal}{Phys. Rev. B}
  \textbf{\bibinfo{volume}{70}}, \bibinfo{pages}{205116}
  (\bibinfo{year}{2004}).

\bibitem[{\citenamefont{Poteryaev et~al.}(2007)\citenamefont{Poteryaev,
  Tomczak, Biermann, Georges, Lichtenstein, Rubtsov, Saha-Dasgupta, and
  Andersen}}]{Poteryaev:prb07}
\bibinfo{author}{\bibfnamefont{A.~I.} \bibnamefont{Poteryaev}},
  \bibinfo{author}{\bibfnamefont{J.~M.} \bibnamefont{Tomczak}},
  \bibinfo{author}{\bibfnamefont{S.}~\bibnamefont{Biermann}},
  \bibinfo{author}{\bibfnamefont{A.}~\bibnamefont{Georges}},
  \bibinfo{author}{\bibfnamefont{A.~I.} \bibnamefont{Lichtenstein}},
  \bibinfo{author}{\bibfnamefont{A.~N.} \bibnamefont{Rubtsov}},
  \bibinfo{author}{\bibfnamefont{T.}~\bibnamefont{Saha-Dasgupta}},
  \bibnamefont{and} \bibinfo{author}{\bibfnamefont{O.~K.}
  \bibnamefont{Andersen}}, \bibinfo{journal}{Phys. Rev. B}
  \textbf{\bibinfo{volume}{76}}, \bibinfo{pages}{085127}
  (\bibinfo{year}{2007}).

\bibitem[{Foo({\natexlab{b}})}]{Footnote2}
\bibinfo{journal}{Let us note that if we do not take the full frequency dependence
  of the Coulomb interaction into account, or, alternatively, the Bose ansatz
  renormalization \cite{Casula:prl12,Tomczak:epl12}, we consider the larger constrained LDA
  values of $U'$ more appropriate; these also give a better agreement between
  DFT+DMFT and experiment \cite{Taranto:prb13}. The estimated 5$\%$ increase of
  the interaction strength due to a reduced screening, however reflects in the
  grey shaded region of realisitc interaction strengths in Fig. \ref{Fig2}.}

\bibitem[{\citenamefont{Taranto et~al.}(2013)\citenamefont{Taranto, Kaltak,
  Parragh, Sangiovanni, Kresse, Toschi, and Held}}]{Taranto:prb13}
\bibinfo{author}{\bibfnamefont{C.}~\bibnamefont{Taranto}},
  \bibinfo{author}{\bibfnamefont{M.}~\bibnamefont{Kaltak}},
  \bibinfo{author}{\bibfnamefont{N.}~\bibnamefont{Parragh}},
  \bibinfo{author}{\bibfnamefont{G.}~\bibnamefont{Sangiovanni}},
  \bibinfo{author}{\bibfnamefont{G.}~\bibnamefont{Kresse}},
  \bibinfo{author}{\bibfnamefont{A.}~\bibnamefont{Toschi}}, \bibnamefont{and}
  \bibinfo{author}{\bibfnamefont{K.}~\bibnamefont{Held}},
  \bibinfo{journal}{Phys. Rev. B} \textbf{\bibinfo{volume}{88}},
  \bibinfo{pages}{165119} (\bibinfo{year}{2013}).

 \bibitem[{\citenamefont{Laverock et~al.}(2013)\citenamefont{Laverock, Chen, 
 Smith, Singh, Balakrishnan, Gu, Lu, Wolf, Qiao, Yang
  et~al.}}]{Laverock:prl13}
\bibinfo{author}{\bibfnamefont{J.}~\bibnamefont{Laverock}},
  \bibinfo{author}{\bibfnamefont{B.}~\bibnamefont{Chen}},
  \bibinfo{author}{\bibfnamefont{K.~E.} \bibnamefont{Smith}},
  \bibinfo{author}{\bibfnamefont{R.~P.} \bibnamefont{Singh}},
  \bibinfo{author}{\bibfnamefont{G.}~\bibnamefont{Balakrishnan}},
  \bibinfo{author}{\bibfnamefont{M.}~\bibnamefont{Gu}},
  \bibinfo{author}{\bibfnamefont{J.~W.} \bibnamefont{Lu}},
  \bibinfo{author}{\bibfnamefont{S.~A.} \bibnamefont{Wolf}},
  \bibinfo{author}{\bibfnamefont{R.~M.} \bibnamefont{Qiao}},
  \bibinfo{author}{\bibfnamefont{W.}~\bibnamefont{Yang}}, 
\bibnamefont{et~al.},
  \bibinfo{journal}{Phys. Rev. Lett.} \textbf{\bibinfo{volume}{111}},
  \bibinfo{pages}{047402} (\bibinfo{year}{2013}).

\bibitem[{\citenamefont{Nakano et~al.}(2012)\citenamefont{Nakano, Shibuya,
  Okuyama, Hatano, Ono, Kawasaki, Iwasa, and Tokura}}]{Nakano:nat12}
\bibinfo{author}{\bibfnamefont{M.}~\bibnamefont{Nakano}},
  \bibinfo{author}{\bibfnamefont{K.}~\bibnamefont{Shibuya}},
  \bibinfo{author}{\bibfnamefont{D.}~\bibnamefont{Okuyama}},
  \bibinfo{author}{\bibfnamefont{T.}~\bibnamefont{Hatano}},
  \bibinfo{author}{\bibfnamefont{S.}~\bibnamefont{Ono}},
  \bibinfo{author}{\bibfnamefont{M.}~\bibnamefont{Kawasaki}},
  \bibinfo{author}{\bibfnamefont{Y.}~\bibnamefont{Iwasa}}, \bibnamefont{and}
  \bibinfo{author}{\bibfnamefont{Y.}~\bibnamefont{Tokura}},
  \bibinfo{journal}{Nature} \textbf{\bibinfo{volume}{487}},
  \bibinfo{pages}{459} (\bibinfo{year}{2012}).

\bibitem[{\citenamefont{Jeong et~al.}(2013)\citenamefont{Jeong, Aetukuri, Graf,
  Schladt, Samant, and Parkin}}]{Jeong:sci13}
\bibinfo{author}{\bibfnamefont{J.}~\bibnamefont{Jeong}},
  \bibinfo{author}{\bibfnamefont{N.}~\bibnamefont{Aetukuri}},
  \bibinfo{author}{\bibfnamefont{T.}~\bibnamefont{Graf}},
  \bibinfo{author}{\bibfnamefont{T.~D.} \bibnamefont{Schladt}},
  \bibinfo{author}{\bibfnamefont{M.~G.} \bibnamefont{Samant}},
  \bibnamefont{and} \bibinfo{author}{\bibfnamefont{S.~S.~P.}
  \bibnamefont{Parkin}}, \bibinfo{journal}{Science}
  \textbf{\bibinfo{volume}{339}}, \bibinfo{pages}{1402} (\bibinfo{year}{2013}).

\bibitem{footnote3}  For thermochromic applications of VO2, see
J. M. Tomczak and S. Biermann, Phys. Status Solidi (b)  {\bf 246},1996 (2009).

\bibitem[{\citenamefont{Tomczak et~al.}(2012)\citenamefont{Tomczak, Casula,
  Miyake, Aryasetiawan, and Biermann}}]{Tomczak:epl12}
\bibinfo{author}{\bibfnamefont{J.~M.} \bibnamefont{Tomczak}},
  \bibinfo{author}{\bibfnamefont{M.}~\bibnamefont{Casula}},
  \bibinfo{author}{\bibfnamefont{T.}~\bibnamefont{Miyake}},
  \bibinfo{author}{\bibfnamefont{F.}~\bibnamefont{Aryasetiawan}},
  \bibnamefont{and} \bibinfo{author}{\bibfnamefont{S.}~\bibnamefont{Biermann}},
  \bibinfo{journal}{EPL (Europhysics Letters)} \textbf{\bibinfo{volume}{100}},
  \bibinfo{pages}{67001} (\bibinfo{year}{2012}).

\bibitem[{\citenamefont{Casula et~al.}(2012)\citenamefont{Casula, Werner,
  Vaugier, Aryasetiawan, Miyake, Millis, and Biermann}}]{Casula:prl12}
\bibinfo{author}{\bibfnamefont{M.}~\bibnamefont{Casula}},
  \bibinfo{author}{\bibfnamefont{P.}~\bibnamefont{Werner}},
  \bibinfo{author}{\bibfnamefont{L.}~\bibnamefont{Vaugier}},
  \bibinfo{author}{\bibfnamefont{F.}~\bibnamefont{Aryasetiawan}},
  \bibinfo{author}{\bibfnamefont{T.}~\bibnamefont{Miyake}},
  \bibinfo{author}{\bibfnamefont{A.~J.} \bibnamefont{Millis}},
  \bibnamefont{and} \bibinfo{author}{\bibfnamefont{S.}~\bibnamefont{Biermann}},
  \bibinfo{journal}{Phys. Rev. Lett.} \textbf{\bibinfo{volume}{109}},
  \bibinfo{pages}{126408} (\bibinfo{year}{2012}).

\end{thebibliography}

\end{document}